\documentclass[twocolumn,showpacs,preprintnumbers,amsmath,amssymb]{revtex4}

\usepackage{graphicx}% Include figure files
\usepackage{dcolumn}% Align table columns on decimal point
\usepackage{bm}% bold math
\usepackage{psfig}

\begin{document}

%\preprint{   }

\title{Spin-dependent effective interactions for halo nuclei}

\author{E. Garrido} 
\email{imteg57@pinar2.csic.es}
\affiliation{ Instituto de Estructura de la Materia, CSIC, 
Serrano 123, E-28006 Madrid, Spain }

\author{D.V. Fedorov}
\author{A.S. Jensen}
\affiliation{ Department of Physics and Astronomy,
        University of Aarhus, DK-8000 Aarhus C, Denmark }

\date{\today}

\begin{abstract}
We discuss the spin-dependence of the effective two-body interactions
appropriate for three-body computations.  The only reasonable choice
seems to be the fine and hyperfine interactions known for atomic
electrons interacting with the nucleus.  One exception is the
nucleon-nucleon interaction imposing a different type of symmetry.  We
use the two-neutron halo nucleus $^{11}$Li as illustration.  We
demonstrate that models with the wrong spin-dependence are basically
without predictive power. The Pauli forbidden core and valence states must 
be consistently treated.
\end{abstract}


\maketitle

\paragraph*{Introduction.}

Quantum halos occur in several branches of physics although mostly
discussed in molecules and nuclei \cite{han95,esr96,nie01,abd02}.
These structures are described in terms of a few weakly bound
composite clusters. The corresponding two-body interactions should in
principle be derived from the basic interactions between the particles
within the clusters.  However, to be practical effective forces must
be employed for at least two reasons. First the calculations simplify
and second the inherent inaccuracy in the calculations from first
principles is often much worse than in the few-body computations.  Thus
the accuracy and inter-dependence of computed observables can be
drastically improved by use of phenomenological effective
interactions.  This division between understanding the basic
interactions and understanding the resulting many-body system is
strikingly illustrated by nuclear structure which was described
virtually independent from detailed knowledge of the nucleon-nucleon
interaction
\cite{boh69}.

In descriptions of nuclear halos effective two-body interactions are
indeed always constructed and used with only superficial connection to
the fundamental forces \cite{ber91,zhu93,tho94,fed95,vin96,esb97}. 
This strategy is
particularly well suited for the spatially extended halos where
details of the potentials are unimportant.  Only low-energy scattering
properties are crucial for the gross structures.  However, more
quantities are both computed and measured with ever increasing
accuracy \cite{gri00,gar02b,mei02,chr02,san03,ney03}. At some point we
are bound to exceed the validity limit of the models.  Designing the
models to maximum performance is very desirable. Their range of
applicability can then be extended by careful choices of model space
and form of the effective interactions.

Few-body computations with spin-dependent effective interactions have
so far not been very abundant \cite{ban92,tho94,fed95,gri03} but this
is likely to change in the near future. It is then important a
correct starting point which unavoidably is the two-body interaction.
The purpose of this report is to discuss the constraints on the
spin-dependence of the effective two-body interactions.  We shall test
by application to three-body nuclear halo systems where the effects
are more indirect, to some extent even hidden, and only revealed by
systematic computations.

\paragraph*{Symmetry requirements.}

It was noted early in the history of nuclear physics that a
spin-dependence in the nucleon mean-field potential was necessary to
reproduce the magic numbers \cite{boh69}. The only rotationally
invariant, parity conserving, spin-dependent quantity is the
spin-orbit potential which is doing the job and therefore used in all
realistic mean-field computations. In complete analogy the
nucleon-core effective interaction for zero core-spin can then only
have central and spin-orbit terms.

On the other hand the nucleon-nucleon interaction has central,
spin-spin, spin-orbit, quadratic spin-orbit, and tensor terms
\cite{mac01}. For an even-even nucleus they reduce to central and 
spin-orbit terms after applying the mean-field average as in all
optical model computations \cite{gre69}.  However, for non-zero
core-spin all these spin-dependent forms are allowed in the
construction of the nucleon-core effective potential.  We shall in
this report for simplicity omit quadratic spin-orbit and tensor
interactions.  The first is usually small and the effect of the second
is reproduced by an orbital angular momentum dependent central force
if the non-diagonal part is neglected or does not contribute.  In
few-body physics this is consistent with the common choice of Hilbert
space with only a few opposite parity relative states, e.g. $s$ and
$p$-states.

This leaves us with central, spin-spin and spin-orbit terms. The
spin-dependence of the effective two-body interaction must then be a
combination of the three possible independent scalar quantities, i.e.
\begin{equation}  \label{eq1}
a \bm{s}_n \cdot \bm{s}_c + b \bm{\ell} \cdot \bm{s}_n
 + c \bm{\ell} \cdot \bm{s}_c \;,
\end{equation}
where $\bm{s}_n$ and $\bm{s}_c$ are the spins of the two particles,
$\bm{\ell}$ is the relative orbital angular momentum operator and
$a,b,c$ are constants. The spin-symmetric combination,
\begin{equation} \label{eq2}
 a \bm{s_n} \cdot \bm{s}_c + b \bm{\ell} \cdot (\bm{s}_n +  \bm{s}_c) \;,
\end{equation}
has to be used for identical particles like nucleons \cite{mac01}.
However, an asymmetric system like the electron and the nucleus
require the combinations
\begin{equation} \label{eq5}
a (\bm{\ell} + \bm{s_n}) \cdot \bm{s}_c + b \bm{\ell} \cdot \bm{s}_n  \;,
\end{equation}
where ``n'' now refers to the electron and $\bm{s}_c$ is the total
angular momentum of the nucleus. The origin of the first term in
eq.(\ref{eq5}) is the magnetic hyperfine interaction arising from the 
nuclear and electron spins \cite{wei78}.

Other combinations are clearly also possible, e.g.
\begin{eqnarray} \label{eq4} 
a \bm{s_n} \cdot \bm{s}_c &+& b \bm{\ell} \cdot \bm{s}_n  \; , \\
(\bm{\ell} + \bm{s_n}) \cdot (\bm{\ell} + \bm{s}_c) &=& 
 \bm{\ell} \cdot \bm{\ell}  + \bm{s}_n \cdot \bm{s}_c +
\bm{\ell}\cdot (\bm{s}_n + \bm{s}_c) \; , \label{eq2a} 
\end{eqnarray}
where eq.(\ref{eq2a}) is more symmetric but equivalent to
eq.(\ref{eq2}) with the $\bm{\ell} \cdot \bm{\ell}$ term included in
the central potential.

For the nucleon-core interaction the absence of symmetry requirements
is complicating the choice. One option could be to use eq.(\ref{eq2})
as in \cite{gar97b,gri03}, where three-body systems with two nucleons
outside a core of finite spin are investigated.  With appropriate
parameter adjustments the computed spectra obtained in
\cite{gri03} and the results involving both the ground state structure and
breakup reactions are indeed quite reasonable \cite{gar97b}.

When the combinations in eq.(\ref{eq2}) are used the total spin
$\bm{s}=\bm{s}_c+\bm{s}_n$ of the neutron-core system is a conserved
quantum number. When the core has a finite spin the two values of
$s=s_c \pm 1/2$ decouple and separate through the spin-spin term. Each
of these two spin values is then split up by the spin-orbit term
according to the possible total two-body angular momenta
$\bm{j}^2=(\bm{\ell}+\bm{s})^2$.  However $s$ and $j$ are not the
usual mean-field quantum numbers, where every nucleon moves in an
orbit characterized by the relative nucleon-core orbital $\ell$ and
nucleon total $j_n=\ell \pm 1/2$ angular momentum. Instead $j_n$
couples to $s_c$ to give the total two-body angular momentum $j$.

One serious problem with eq.(\ref{eq2}) arises because the nucleon
angular momentum $j_n$ is not conserved like the total spin.
Therefore the usual mean-field spin-orbit partners with $j_n=\ell \pm
1/2$ are inevitably mixed in both two and three-body systems. The
essence of the difficulty is that the motion of the valence nucleon(s)
outside the core is inconsistent with the (approximate mean-field)
motion of the identical nucleons within the core. This is obviously a
problem for a few-body description in terms of components with given
$j_n$ as needed when we want to bridge the gap to mean-field
computations.

The problem is much worse if one and only one of the mixed spin-orbit
partners is Pauli forbidden by core occupation as the $p_{3/2}$ orbit
in $^{10}$Li and $^{11}$Li.  This may be disguised by parameter
adjustments, but sufficiently many independent observables computed
from the same parameter set would undoubtedly reveal the problem.
Computations of relatively few observables may be quite reasonable
although uncontrolled and thus suspicious.

We need to restore the mean-field single-particle total angular
momentum as a conserved quantum number.  Therefore in subsequent
calculations (see for example \cite{gar99b}) the spin-orbit
interaction was changed and the spin-dependence were instead given by
eq.(\ref{eq4}).  Now the dominating spin-orbit term conserves $\ell$
and ${j_n}$ as indicated by the usual notation $\ell_{j_n}$.  The
spin-spin term should then be a perturbation precisely designed to
split the states with different total angular momenta arising from
couplings of core ($s_c$) and nucleon ($j_n$) spins.  This is a
reasonable assumption when the single particle spin-orbit splitting is
much larger and essentially is maintained.

To achieve full consistency with the mean-field description we have to
replace eq.(\ref{eq2}) with eq.(\ref{eq5}), which conserves ${j}_n$.
It is interesting that this choice is appropriate for nucleon-core as
well as for for electron-nucleus interactions. In both cases the fine-
and hyperfine structure can apparently then be properly described as
originating from the spin-spin and spin-orbit terms, respectively.
The difference between eqs.(\ref{eq5}) and (\ref{eq4}) is marginal for
the present cases of interest, since the strength of the spin-spin
term in both cases must be rather small. However, eq.(\ref{eq5}) is
clearly preferable from a conceptual point of view where the
connection to mean-field computations is necessary or at least very
desirable.

It is worth emphasizing that eqs.(\ref{eq2}), (\ref{eq5}),
({\ref{eq4}) and (\ref{eq2a}) are identical both for $s$-waves and for
zero core spin.

\paragraph*{Comparing two-body properties.}

The interaction must in general be combinations of three terms as in
eq.(\ref{eq1}) with appropriate individual radial form factors. The
relative importance of the terms in eq.(\ref{eq2}) is illustrated for
$d$-waves in the upper part of fig.\ref{fig1}. We define $a=x$ and
$b=\sqrt{1-x^2}$ such that the parameter $x$ controls the relative
weight between the spin-spin and spin-orbit parts of these potentials.
For $x=\pm 1$ only the spin-spin term enters, the total spin is
conserved, and the state with $s=0$ is separated from the three
degenerated $s=1$ states.  When $|x|<1$ the spin-orbit part breaks
this degeneracy leaving four different energies as for $x=0$ where
only the spin-orbit term contributes.

\begin{figure}
\includegraphics[scale=0.4,angle=-90]{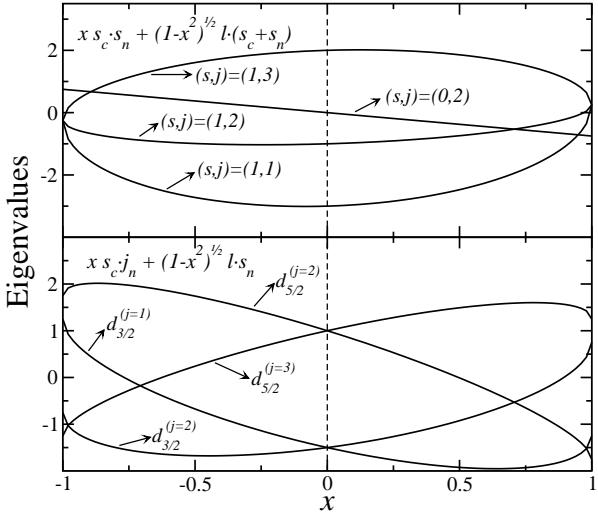}
\caption{\label{fig1} Eigenvalues for $d$-waves and $s_c=s_n=1/2$ 
of the potentials in eqs.(\ref{eq2}) (upper part) and (\ref{eq5})
(lower part) as functions of $x$ where $a=x$ and $b=\sqrt{1-x^2}$.  We
use the notation $\ell_{j_n}^{(j)}$. The orbital $(\ell)$, nucleon
$({j_n})$ and total $(j)$ angular momenta are defined in the text.}
\end{figure}

The relative importance of the terms in eq.(\ref{eq5}) is illustrated
in the lower part of fig.\ref{fig1}, where we again use the parameter
$x$ as a measure of the relative weights of spin-spin and spin-orbit
potentials.  For $x=0$ (only spin-orbit) the eigenvalues coincide for
the same value of $j_n$.  For finite $x$ the spin-spin interaction
removes this degeneracy leaving four different eigenvalues
corresponding to the different pairs of $j_n$ and $j$.

The difference between the interactions in eqs.(\ref{eq2}) and
(\ref{eq5}) is illustrated in fig.\ref{fig2} with the potential
\begin{equation}
V(x)=\bm{s}_n\cdot\bm{s}_c + B \bm{\ell} \cdot (\bm{s}_n+\bm{s}_c)+
x (1-B) \bm{\ell}\cdot \bm{s}_c \;,
\label{eq7}
\end{equation}
varying from the form in eq.(\ref{eq2}) for $x=0$ to that of
eq.(\ref{eq5}) for $x=1$.  In both cases $B$ is the strength of the
spin-orbit term.  We have chosen a large value of $B$ to enhance the
effect of the spin-orbit term. For $x=0$ the total spin $s$ and total
angular momentum $j$ are conserved quantum numbers collectively
denoted by $s_j$.  When $x$ increases $s$ and $j$ are not conserved
quantum numbers anymore, and the eigenvalues evolve as shown in
fig.\ref{fig2}. For $x=1$ when we have the form in eq.(\ref{eq5}) the
conserved quantum numbers $(\ell,j_n,j)$ are collected in the notation
$\ell_{j_n}^{(j)}$.

Let us first focus on the upper part of the figure. For $x=1$ we see
clearly the spin-orbit splitting of the $d_{5/2}$ and
$d_{3/2}$-states.  The $d_{5/2}$-states are lower due to the negative
sign of the strength $B$. This illustrates that the choice in
eq.(\ref{eq5}) ($x=1$) permits construction of low-lying
$d_{5/2}$-states without contribution from the $d_{3/2}$.  This is not
possible with eq.(\ref{eq2}) ($x=0$), where two of the eigenvalues mix
the $d_{5/2}$ and $d_{3/2}$-states, while the other two are pure
$d_{3/2}$ and pure $d_{5/2}$, respectively.  Thus it is not possible
with the choice in eq.(\ref{eq2}) to have two low-lying
$d_{5/2}$-states without admixtures from the $d_{3/2}$-state.

\begin{figure}
\includegraphics[scale=0.4,height=9.5cm,width=9.0cm,angle=-90]{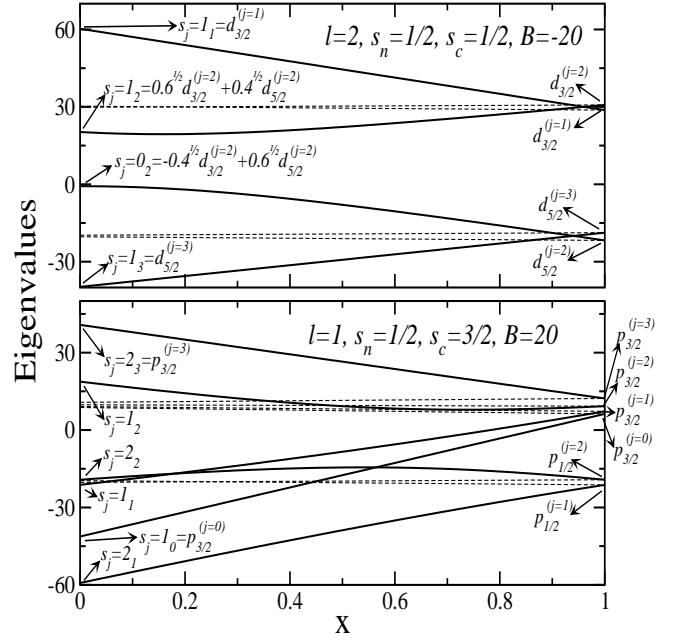}
\caption{\label{fig2} Eigenvalues of the potentials eqs.(\ref{eq7})
(solid lines) and (\ref{eq8}) (dashed lines). The upper part is for
$d$-waves and $s_c$=$s_n$=1/2, and the lower part is for $p$-waves and
$s_c$=3/2,$s_n$=1/2. These are the waves relevant for $^{17}$Ne and
$^{11}$Li, respectively.  When conserved, the total spin $s$ and total
$j$ angular momentum are collected in the label $s_j$
and analogously $\ell_{j_n}^{(j)}$ labels state with conserved
orbital, particle ``n'' and total $j$ angular momentum.  $B$ is the
strength of the spin-orbit interaction.}
\end{figure}

Similarly, the dashed lines in fig.\ref{fig2} are the eigenvalues of
the potential
\begin{equation}
V(x)=\bm{s}_n\cdot\bm{s}_c + B \bm{\ell} \cdot \bm{s}_n +
x \bm{\ell}\cdot \bm{s}_c
\label{eq8}
\end{equation}
that for $x=0$ and $1$ reduce to eqs.(\ref{eq4}) and (\ref{eq5}),
respectively.  We see that although $j_n$ is not a conserved quantum
number even for $x=0$, the eigenvalues remain almost constant all the
way up to $x=1$.  This result is a consequence of the small absolute
size of the spin-spin interaction compared to the spin-orbit
term. Under this realistic condition (\ref{eq4}) is a good
approximation to (\ref{eq5}).

In the lower part of fig.\ref{fig2} we show the same kind of results
for $p$-waves and $s_c$=3/2. We start again with the expression
eq.(\ref{eq7}) but now with $B=20$. The positive value of $B$ reverses
the normal order of the spin-orbit splitting. Then for $x$=1, where
$j_n$ and $j$ are good quantum numbers, the eigenvalues with smallest
value of $j_n$ are the lowest.  This is a useful tool to exclude
occupation of the Pauli forbidden $p_{3/2}$-states achieved by
shifting them to high energies.  Again we see that eq.(\ref{eq5})
($x$=1) allows close lying, still spin-split, states with equal value
of $j_n$.  In contrast eq.(\ref{eq2}) ($x=0$) corresponds to two pure
$j_n$=3/2-states and four other states which mix the $p_{1/2}$ and the
$p_{3/2}$ orbits according to
\begin{subequations}
\begin{eqnarray}
s_j=1_1&=&-\sqrt{1/6}p_{1/2}^{(j=1)}+\sqrt{5/6}p_{3/2}^{(j=1)}, \label{eq8a}\\
s_j=1_2&=&-\sqrt{1/2}p_{1/2}^{(j=2)}+\sqrt{1/2}p_{3/2}^{(j=2)}, \\
s_j=2_1&=& \sqrt{5/6}p_{1/2}^{(j=1)}+\sqrt{1/6}p_{3/2}^{(j=1)}, \\
s_j=2_2&=& \sqrt{1/2}p_{1/2}^{(j=2)}+\sqrt{1/2}p_{3/2}^{(j=2)}. \label{eq8d} 
\end{eqnarray}
\end{subequations}

It is then clear that when eq.(\ref{eq2}) is used, the two low-lying
$p$-states necessarily contain part of the $p_{3/2}$-states. They are
forbidden by the Pauli principle for both $^{10}$Li and $^{11}$Li. The
dashed lines in the lower part of fig.\ref{fig2} show eigenvalues
obtained from the potential eq.(\ref{eq8}). Again we observe that
eq.(\ref{eq4}) ($x=0$) is a good approximation to eq.(\ref{eq5})
($x=1$) provided the spin-spin term only is a perturbation to the
spin-orbit term.

\paragraph*{Comparing three-body properties.}

Let us now investigate how the different choices of the nucleon-core
interaction affect the full three-body calculation.  We take
$^{11}$Li ($^9$Li+$n$+$n$) as an example with the Hilbert space
consisting of $s$ and $p$-waves. Then $s_c$=3/2 and the lowest neutron
$s_{1/2}$ and $p_{3/2}$-orbits are fully occupied in the $^9$Li core,
and therefore not available for the valence neutrons due to the Pauli
principle.  The calculations are performed by solving the Faddeev
equations with the hyperspheric adiabatic expansion method
\cite{gar97}. For the neutron-neutron interaction we use 
eq.(\ref{eq2}) supplemented by a central part. The radial form factors
are gaussians, and their parameters are adjusted to reproduce low
energy nucleon-nucleon scattering data.

For the nucleon-core interaction we first use eq.(\ref{eq5}) plus a
central part, i.e.
\begin{equation}
V_{nc}^{(\ell)}=V_c^{(\ell)}+V_{ss}^{(\ell)} \bm{s}_c \cdot \bm{j}_n
               +V_{so}^{(\ell)} \bm{\ell} \cdot \bm{s}_n
\label{eq10}
\end{equation}
We use gaussian radial form factors adjusted independently for $s$ and
$p$-waves. To account for the Pauli principle for the $s_{1/2}$ and
$p_{3/2}$ states we start by using a simple shallow $s$-wave interaction
without bound states and a large repulsive spin-orbit strength
shifting the $p_{3/2}$-states to high energy as seen in the lower part
of fig.\ref{fig2} for $x$=1.

As in \cite{gar97} we choose a range for the $^9$Li-neutron
interaction of 2.55 fm, and adjust the strengths of $V_c^{(\ell=1)}$
and $V_{ss}^{(\ell=1)}$ to place the two low-lying spin-split
$p_{1/2}$-resonances in $^{10}$Li at 0.3 and 0.5 MeV computed as poles
of the $S$-matrix. These values are consistent with the available
experimental data \cite{gar02}.  We finally use the strength of the
central $s$-wave interaction to fit the experimental $^{11}$Li 
separation energy, and the strength of the $s$-wave spin-spin
potential to place a low-lying virtual $s$-state in $^{10}$Li 
at 50 keV, as indicated by experiments. 

The precise values of the parameters in the potentials are given in
the caption to fig.\ref{fig3}. The computed r.m.s. radius is 3.3 fm,
and the $p$-wave content is slightly above 30\%. Use of a weaker central
$s$-wave interaction can increase
the $p$-wave content up to around 40\% in somewhat better agreement 
with the known experimental values. In this case an additional effective
three-body potential is needed to maintain the two-neutron separation
energy at the right value. However, this fine tuning is not 
necessary for the present investigation. The $^{10}$Li energy distribution is
approximately obtained as the Fourier transform
$\Psi(\bm{p}_x,\bm{p}_y)$ of the $^{11}$Li wave function, i.e.
\begin{equation}
\frac{dn}{dE_x}=
\mu_x p_x \int |\Psi(\bm{p}_x,\bm{p}_y)|^2 d^3p_y d\Omega_{p_x}
\label{eq10b}
\end{equation}
where $\bm{p}_x$ is the $^9$Li-neutron relative momentum whose
direction is given by $\Omega_{p_x}$, $\mu_x$ is the $^9$Li-neutron
reduced mass, $\bm{p}_y$ is the relative momentum between the second
neutron and the $^{10}$Li center of mass, and $E_x=p_x^2/2\mu_x$.

\begin{figure}
\vspace*{-1.2cm}
\includegraphics[scale=0.45,angle=-90]{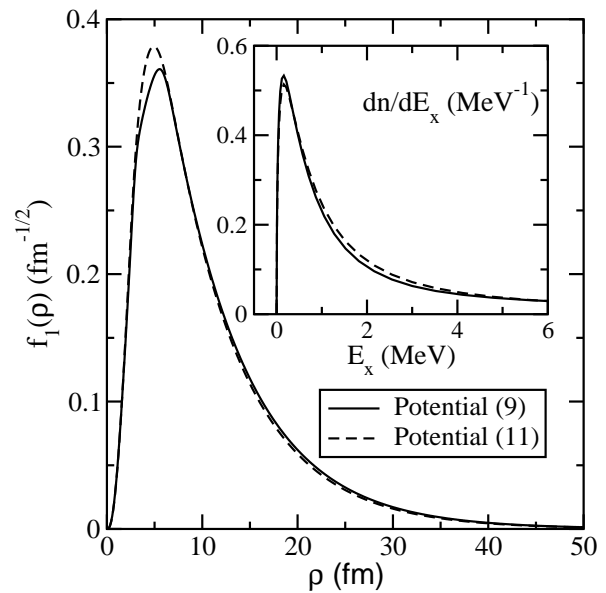}
\vspace*{-0.7cm}
\caption{\label{fig3} The dominating radial component for the ground 
state of $^{11}$Li (outer part) and the $^{10}$Li energy distribution
as given in eq.(\ref{eq10b}) (inner part). The potentials in
eqs.(\ref{eq10}) (solid line) and (\ref{eq11}) (dashed line) are used.
The strengths of the gaussians $V_c^{(\ell=0)}$, $V_{ss}^{(\ell=0)}$,
$V_c^{(\ell=1)}$, $V_{ss}^{(\ell=1)}$, and $V_{so}^{(\ell=1)}$ are
$-7.05$ MeV, $-1.6$ MeV, 260.25 MeV, 1.0 MeV and 300 MeV,
respectively, for the potential eq.(\ref{eq10}) and $-5.95$ MeV,
$-1.6$ MeV, $-27.125$ MeV, 1.5 MeV and 5 MeV for the potential
eq.(\ref{eq11}). The range of the gaussians is 2.55 fm in all the
cases.}
\end{figure}

Let us carry out the same calculations but now using eq.(\ref{eq2}).
The $^9$Li-neutron interaction is given as
\begin{equation}
V_{nc}^{(\ell)}=V_c^{(\ell)}+V_{ss}^{(\ell)} \bm{s}_n \cdot \bm{s}_c
               +V_{so}^{(\ell)} \bm{\ell} \cdot (\bm{s}_n+\bm{s}_c).
\label{eq11}
\end{equation}
The $p$-resonances are now characterized by the quantum numbers given
in the lower part of fig.\ref{fig2} for $x$=0.  We proceed as before,
placing the two lowest $p$-resonances at the energies of 0.3 MeV and
0.5 MeV and using the spin-orbit strength to shift the remaining four
states to higher energies.  Again the $s$-wave interaction is used to
adjust to the experimental separation energy of 0.3 MeV.  This
procedure results in a r.m.s. radius of 3.4 fm, also consistent with
the experimental data, but the $p$-wave content is now too small, less
than 7\%.

An increase of this $p$-wave content without changing the energies of the
two lowest resonances can only be achieved by reducing the strength of
the spin-orbit interaction.  With the parameters given in the caption
of fig.\ref{fig3} we then obtain a reasonable $^{11}$Li wave function,
with the right separation energy, r.m.s. radius, and a $p$-wave
content of 27\%.  Furthermore, as shown by the dashed curves in
fig.\ref{fig3} the dominating radial wave function as well as the
$^{10}$Li energy distribution obtained with the potential
eq.(\ref{eq11}) resembles those obtained with eq.(\ref{eq10}).

However, some hidden differences remain.  First, the
$p_{3/2}$-occupation probability only amounts to 7\% with the
potential in eq.(\ref{eq10}) and around 13\%, more than half of the
total $p$-wave contribution, with the potential eq.(\ref{eq11}).
Second, by increasing the repulsion in the spin-orbit interaction in
potential eq.(\ref{eq10}) this $p_{3/2}$-probability can be reduced
without significant changes of the computed $^{11}$Li properties.
However, with potential eq.(\ref{eq11}) it is not possible to reduce
the $p_{3/2}$-probability and still keep the $p$-wave content at a
realistic fairly high value.  Third with the potential in
eq.(\ref{eq10}) the computed $^{10}$Li has only two low-lying
resonances of $p_{1/2}$ character, while for eq.(\ref{eq11}) almost
all the $p$-resonances are rather close-lying, i.e. the five lowest
energies (computed as poles of the $S$-matrix) are 0.3, 0.5, 1.0, 1.3
and 1.6 MeV.

Let us now maintain the realistic properties of the $^{11}$Li ground
state and turn to the lowest $1^-$ excitation, i.e. the $1/2^+$ state.
The only difference in the calculation is that the inclusion of the
Faddeev components must be consistent with the new quantum
numbers. The same potentials as for the ground state must be used,
since precisely the same partial waves constitute both ground state
and excited state configurations.

Now drastic differences appear.  For the potential in eq.(\ref{eq10})
the computed $1/2^+$ state is unbound.  Furthermore, a calculation
with the complex scaling method reveals a $1/2^+$ resonance at 0.4 MeV
with a width of 0.3 MeV.  These values are consistent with the
detailed calculations shown in \cite{gar02b}. However, when the
potential in eq.(\ref{eq11}) is used the $1/2^+$ state is bound by 1.3
MeV, i.e. even more bound than the ground state.  We can then conclude
that when only ground state properties are used to fit the potential
parameters, the excited states are not automatically also correct even
when they are spanned by the same Hilbert space.

For the $1^-$ excitations of $^{11}$Li basically one $s$ and one
$p$-state are simultaneously occupied unlike the ground state where
both neutrons simultaneously are located in pairs of either $s$ or
$p$-states.  The 1/2$^+$ resonance obtained with eq.(\ref{eq10}) does
not contain contributions from the $p_{3/2}$-waves when a realistic
spin-orbit splitting is applied.  In contrast this resonance obtained
with eq.(\ref{eq11}) contains a $p_{3/2}$-contribution of about 10\%.
The wrong properties of the interactions may be parameterized away
while still reproducing the ground state observables, but
catastrophically wrong results may return for the other properties for
example of $1^-$ excitations.

\paragraph*{The Pauli principle.}

The calculations shown in the previous section contain an important
simplification concerning the way the Pauli principle is taken into
account. This has been done to make the conclusions clear and avoid
the mixing with technical details that could easily obscure them. We
now shall investigate if a better treatment of the Pauli principle,
necessary for any realistic calculation, is modifying some of the
previous results.  In \cite{gar99} an appropriate method to treat the 
Pauli principle in few-body calculations is described. The realistic
two-body interactions able to bind a nucleon into a Pauli forbidden
state are substituted by the corresponding phase equivalent potentials 
with exactly the same phase shifts but without the Pauli forbidden bound
states.

In \cite{gar02} constraints on the neutron-$^9$Li were derived
consistent with the known $^{10}$Li spectrum and the $^{11}$Li
properties.  In particular it was found that the interaction in
eq.(\ref{eq10}) with gaussian radial form factors with range equal to
2 fm, and strengths for $V_c^{(\ell=0)}$, $V_{ss}^{(\ell=0)}$,
$V_c^{(\ell=1)}$, $V_{ss}^{(\ell=1)}$, and $V_{so}^{(\ell=1)}$ equal
to $-94.0$ MeV, $-11.4$ MeV, $-79.64$ MeV, 1.1 MeV and $-13.12$ MeV,
respectively, results in a $^{10}$Li spectrum with a low-lying 2$^-$
$s_{1/2}$ virtual state at 50 keV, and a 1$^+$/2$^+$ $p_{1/2}$-doublet
at 0.25 MeV and 0.54 MeV, respectively, consistent with the available
experimental data. This neutron-$^9$Li interaction has a deeply bound
$s_{1/2}$ state and a bound $p_{3/2}$ state at $-4.1$ MeV. These two
states are Pauli forbidden and subsequently removed in the three-body
calculation by use of the corresponding phase equivalent potentials
\cite{gar02}.  In this way the $s_{1/2}$ and the $p_{3/2}$ forbidden
states are explicitly excluded from the calculation.

As shown in \cite{gar02}, the use of this two-body interaction
results in a $^{11}$Li ground state with a two-neutron separation
energy of 0.30 MeV, a $p$-wave content of around 40\%, and a r.m.s.
radius of 3.2 fm. These results are obtained after a fine
tuning with an effective three-body interaction that is known to be
necessary in three-body calculations to account for the
polarizations of the particles that are beyond those described by the
two-body interactions. At the same time the $^{11}$Li wave function
obtained is consistent with the experimental invariant mass spectrum,
core momentum distribution and angular distribution obtained after
fragmentation of $^{11}$Li. Furthermore, in \cite{gar02b} we show that
the same interaction gives rise to a $1/2^+$ $^{11}$Li resonance at an
energy above threshold of about 0.6 MeV and a width of 0.5 MeV. These
results are similar to the ones obtained with the calculations in the
previous section for the interaction given in eq.(\ref{eq10}).

Let us perform now the same kind of calculations for potential
(\ref{eq11}). We then use the $s$-wave neutron-$^{9}$Li interaction
specified above, with a deeply bound $s_{1/2}$ state, forbidden by the
Pauli principle, that is removed from the three-body calculation by
use of the phase equivalent potential. For the $p$-wave interaction
the good quantum numbers of the spin operators in the
two-body potential are the total two-body spin $s$ and the total
two-body angular momentum $j$, and therefore the eigenfunctions mix
the $p_{1/2}$ and $p_{3/2}$ states (see eqs.(\ref{eq8a}) to
(\ref{eq8d})). It is then impossible to find an interaction binding
pure $p_{3/2}$ states into a Pauli forbidden state, and with two pure
low-lying $p_{1/2}$ resonances. This is due to the inconsistency
between the good quantum numbers of the interaction (\ref{eq11}) and
the mean-field description used for the core neutrons.

This inconsistency cannot be cured by an appropriate treatment of the
$p$-wave states. Let us nevertheless describe various attempts.
First, we take the same shallow $p$-wave potential as in the previous
section (the strengths of the gaussians of the interaction
(\ref{eq11}) are given in the caption of fig.\ref{fig3}). Proceeding
in this way we observe that an appropriate treatment of the Pauli
principle only in the $s$--waves while keeping unchanged the
$p$--interaction is not giving new results compared to the previous ones.
As before, the $^{11}$Li ground state wave function can be considered 
to be reasonable, but the same interaction
gives rise to a bound $1/2^+$ state with 1 MeV separation energy,
clearly more bound than the experimental ground state.

A second option is to construct an interaction (\ref{eq11}) such that
four of the $p$-wave eigenfunctions are bound, while the other two are
low-lying resonances. The bound states can be interpreted as the
four neutrons occupying the $p$-shell in the $^9$Li core. These four
states should be removed from the calculation, and the
remaining two states would correspond to the $p$-resonances at
0.25 MeV and 0.54 MeV. As mentioned this is in itself
inconsistent, since the excluded states necessarily contain some
$p_{1/2}$ contribution, and the two low-lying $p$-resonances contain
part of the $p_{3/2}$ waves. In any case this can be achieved by
taking gaussian radial potentials in eq.(\ref{eq11}) with a
range of 2 fm and strengths for $V_c^{(\ell=1)}$,
$V_{ss}^{(\ell=1)}$, and $V_{so}^{(\ell=1)}$ equal to $-77.6$ MeV,
$-2.3$ MeV, and $6.8$ MeV.  When this two-body
$p$-potential is used, and the four bound $p$-states are removed from
the three-body calculation, the ground state $^{11}$Li wave function
needs a very large three-body effective interaction to recover the
experimental two-neutron separation energy of 0.3 MeV. Furthermore the
$p$-wave content is extremely large (80\%), and the r.m.s.
radius (2.4 fm) is far below the experimental value. We see
that in this case even the ground state wave function shows important
deficiencies. Also the $1/2^+$ state does not show any low lying
resonance.

A final attempt can be to use a repulsive core only for the $p_{3/2}$
waves. For this purpose we can employ the phase equivalent potential
constructed to be used with interaction (\ref{eq10}). However, this
repulsive potential is implicitly assuming the shell model quantum
numbers for the neutron state, and when used in combination with the
spin operators in eq.(\ref{eq11}) the energies of the different two-body
$p$-states bear no resemblance to the initial $p$-states, and some of
them are even bound.

We have now accounted for the Pauli principle in various ways,
i.e. using a shallow potential without bound states, using phase
equivalent potentials and repulsive cores. The conclusion remains that
the spin-dependent interaction must be simultaneously consistent with
the treatment of both core and valence neutrons. No parameter
variations alter this conclusion.

\paragraph*{Conclusions.}

The first step towards a reliable description of few-body systems is
to construct appropriate two-body interactions between the pairs of
particles. We discuss the spin-dependence of this effective
interaction. Three different terms combining spin and orbital angular
momentum are possible with the necessary rotational invariance and
parity conservation. Various combinations correspond to applications
in different fields of physics. We focus especially on the forms
appropriate for the nucleon-nucleon interaction and the fine and
hyperfine interactions for electrons in the atom.

The properties of different two-body interactions are first
compared. Then the effects on computed three-body structures are
investigated for $^{11}$Li ($n+n+^{9}$Li). We show that parameters
adjusted to reproduce the three-body ground state properties can lead
to wrong results for the excited states depending on the
form of the spin-dependent effective interactions.  We conclude that
only one choice is consistent with the mean-field treatment implicitly
assumed for the core nucleons. Violating this condition compromises
the treatment of the Pauli forbidden states and reliable predictions
are not possible.

\end{document}